\documentclass[showpacs,prd]{revtex4}
\usepackage[dvips]{graphicx}
\usepackage{amsmath}
\newcommand{\Sw}{Schwarzschild }
\begin{document}


\title{Thermodynamical Properties of Horizons}

\author{Jarmo M\"akel\"a} 
\email[Electronic address: ]{jarmo.makela@phys.jyu.fi}  
\author{Ari Peltola}
\email[Electronic address: ]{ari.peltola@phys.jyu.fi} 
\affiliation{Department of Physics, University of Jyv\"askyl\"a, PB 35 (YFL), FIN-40351 Jyv\"askyl\"a, Finland}

\date{May 30, 2002}

\begin{abstract}
We show, by using Regge calculus, that the entropy of any finite part of a Rindler horizon is, in the semi-classical limit, one quarter
of the area of that part. We argue that this result implies that the entropy associated with any horizon of spacetime is, in the semi-classical
limit, one quarter of its area. As an example, we derive the Bekenstein-Hawking entropy law for the \Sw black hole.
\end{abstract}


\maketitle
 
A remarkable discovery was made by Bekenstein and Hawking almost 30 years ago when they found that black hole has
a certain entropy which is one quarter of its horizon area\cite{bek,haw}. As a consequence, black holes emit thermal radiation with a spectrum
which is essentially that of a black body. After Bekenstein and Hawking it was soon discovered by Unruh\cite{unruh,bd,wald} that not 
only black holes
but also the so-called Rindler horizon of an accelerated observer emits thermal radiation. So it appeared that the black hole is
not necessary for the creation -- from the point of view of a certain observer -- of thermal radiation out of the vacuum. Rather,
what is essential for the creation of thermal radiation, is the existence of a \textit{horizon}. In short, it appears that all horizons
of spacetime, no matter whether they are black hole, Rindler, de Sitter, or cosmological horizons, have certain thermodynamical
properties in common. For instance, all horizons of spacetime emit thermal radiation. In particular, it appears -- and
this is one of the main theses of this paper -- that if we consider any part of any spacetime horizon, then one can associate
with that part an entropy which is one quarter of the area of the part.

In this paper we give support to this thesis by means of a detailed analysis of the thermodynamical properties of Rindler horizons. A 
novel feature in our analysis is the use of the \textit{Regge calculus}\cite{regge,mtw,rw} approach to Einstein's general relativity. 
In short, Regge
calculus is an approach to general relativity, where spacetime is modelled by a piecewise flat, or simplicial, manifold and
the geometrial properties of that manifold are coded into its edge lengths. Using Regge calculus we show that in the semi-classical 
limit each
part of the Rindler horizon of spacetime possesses an entropy which is one quarter of its area. Finally, at the end of this
paper, we shall show how the analysis used in the study of Rindler horizons may also be used for the derivation of the
Bekenstein-Hawking law for the black hole entropy.

\begin{figure}[!htb]
\centering
\includegraphics[scale=0.5]{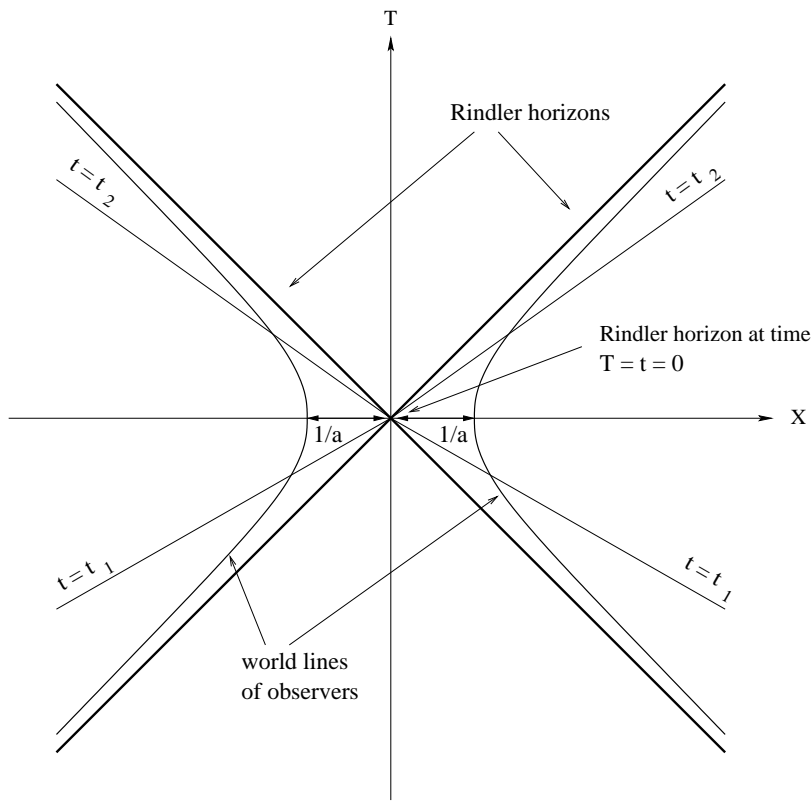}
\caption{Rindler spacetime. The hypersurfaces where $t=$ constant are orthogonal to the world lines of the accelerated observers.
The curvature of these hypersurfaces is concentrated at the Rindler horizon at time $T = 0$.}
\label{fig:masa}
\end{figure}

Rindler horizons are such horizons of spacetime which appear in the rest frame of an accelerated observer. In general, the world 
line of a uniformly accelerated observer in flat two-dimensional Minkowski spacetime has the equation (unless otherwise stated
we shall always have $c=G=\hbar = k_B = 1$)\cite{bd}
\begin{equation} X^2-T^2= \frac{1}{a^2},
\end{equation}
where $a$ is the proper acceleration of the observer, and $X$ and $T$, respectively, are the minkowskian space and time coordinates. 
The world line of the observer may be written in the parametrized form
\begin{equation} \label{eq:tillin} T(t)=\frac{1}{a} \sinh (at),
\end{equation}
\begin{equation} \label{eq:tallin}X(t)=\frac{1}{a} \cosh (at).
\end{equation}
In this expression, $t$ is the proper time of the observer. The world lines of the two observers, being uniformly accelerated in opposite
directions, have been drawn in Fig. \ref{fig:masa}. In that figure we may also see the Rindler horizon (which, in the observer's rest frame is, in 
our two-dimensional figure, represented by a point) of the accelerated observers.

Consider now two oppositely accelerated observers. The spacelike hypersurfaces where the proper times $t$ of the observers are $t_1$
and $t_2$ ($t_1 < t_2$) are orthogonal to the world lines of the accelerated observers. So far we have just recalled the standard 
properties of accelerated observers in flat spacetime. At this point we turn to quantization. We evaluate the propagator
\begin{displaymath} K\big( q_{ab}(t_2),t_2;q_{ab}(t_1),t_1 \big).
\end{displaymath}
This propagator gives the propability amplitude that the metric tensor on the hypersurface $t = t_2$ is $q_{ab}(t_2)$ provided that the
metric tensor on the hypersurface $t=t_1$ was $q_{ab}(t_1)$.
\begin{figure}[!htb]
\centering
\includegraphics[scale=0.5]{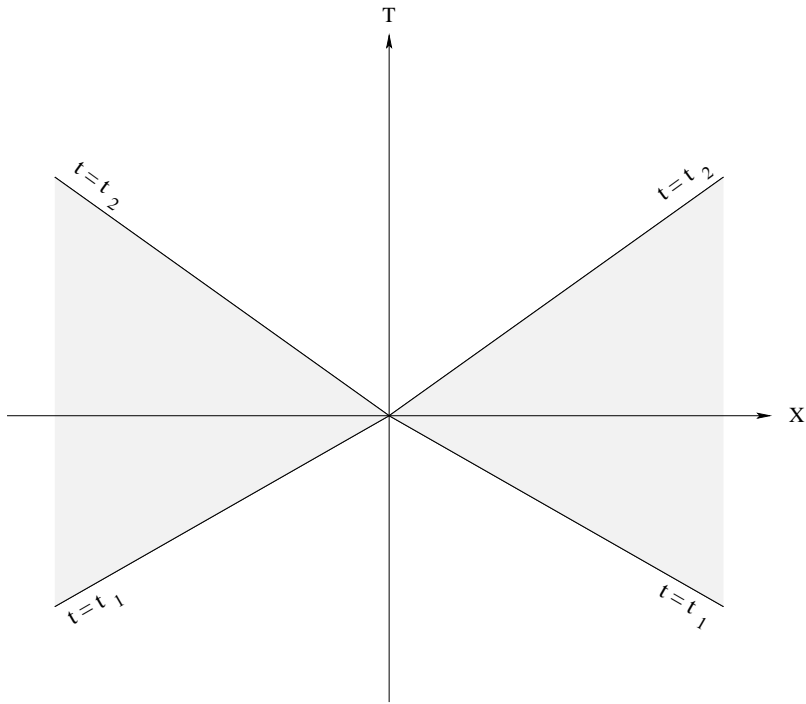}
\caption{The hypersurfaces $t=t_1$ and $t=t_2$ bound a certain region of Rindler spacetime (shaded region in this figure).
When calculating the  propagator we integrate over all four-metrics $g_{\mu \nu}$ in this region. \label{fig:ile}}
\end{figure}
Formally, the propagator may be written as a path integral:
\begin{equation} \label{eq:lee}K\big( q_{ab}(t_2),t_2;q_{ab}(t_1),t_1 \big)= \int \mathcal{D}[g_{\mu \nu}] \exp\big( iS_G[g_{\mu \nu}] \big),
\end{equation}
where the integration has been performed over the four-metrics $g_{\mu \nu}$ between the hypersurfaces $t=t_1$ and $t=t_2$ (shaded
region in Fig. \ref{fig:ile}). $S_G[g_{\mu \nu}]$, in turn, is the classical gravitational action:
\begin{equation} \label{eq:sergei} S_G[g_{\mu \nu}] = \frac{1}{16\pi} \int_M R\sqrt{-g}\, d^4x \, + S_{\partial M},
\end{equation}
where $R$ is the Riemannian scalar and $g$ is the determinant of the four-metric $g_{\mu \nu}$. In the first term on the right
hand side of Eq. (\ref{eq:sergei}) the integration is performed over the spacetime region between the surfaces $t=t_1$ and
$t=t_2$, whereas the  second term is the boundary term to the gravitational action.

At the present state of research it is not known how to evaluate explicitly the right hand side of Eq. (\ref{eq:lee}). However, it is 
possible to approximate
the propagator by means of the so-called semi-classical approximation. In this approximation we may write the propagator as
\begin{equation} K\big( q_{ab}(t_2),t_2;q_{ab}(t_1),t_1 \big) \approx  N\big[ q_{ab}(t_1),q_{ab}(t_2) \big] 
\exp\big( i\widetilde{S}_G [g_{\mu \nu}] \big),
\end{equation}
where $N\big[ q_{ab}(t_1),q_{ab}(t_2) \big]$ is some slowly-varying functional of $q_{ab}(t_1)$ and $q_{ab}(t_2)$, and
$\widetilde{S}_G$ is the gravitational action when Einstein's field equations are satisfied. For flat spacetime $R=0$, and therefore
we are left with the boundary terms only. It is possible to choose the boundary term at asymptotic infinity such that it vanishes
in flat spacetime, and therefore all contribution to the boundary terms comes from the hypersurfaces $t=t_1$ and $t=t_2$.
It is at this point where Regge calculus enters the stage.

In Regge calculus the gravitational action is written as
\begin{equation} S_G= \frac{1}{8\pi} \Big( \sum_j A_j \epsilon_j + \sum_k A_k \phi_k \Big).
\end{equation}
In this expression, $A_j$'s are the areas of those two-simplices, or triangles $j$, which do not lie on the boundary of spacetime, and 
$\epsilon_j$'s are the corresponding deficit angles. $A_k$'s, in turn, are the areas of those triangles $k$ which lie on
the boundary of spacetime, $\phi_k$'s being the corresponding deficit angles. If the triangles $k$ lie on a spacelike hypersurface of
spacetime, then $\phi_k$ may be understood as the boost angle between the timelike normals of the two tetrahedra
having the triangle $k$ in common. If spacetime is flat, we have
\begin{equation} \epsilon_j = 0
\end{equation}
for every $j$, and the gravitational action reduces to
\begin{equation} S_G = \frac{1}{8\pi} \sum_k A_k \phi_k.
\end{equation}
In other words, if we have a spacelike hypersurface embedded in flat spacetime such that its curvature is concentrated along certain
flat two-spaces, we get the action -- up to the term $(8\pi)^{-1}$ -- by just multiplying the areas by the corresponding boost
angles, and summing the products. This result is particularly useful in our present study of Rindler horizons: Rindler horizons
are flat two-spaces, and therefore they may be constructed from flat triangles. Morever, as one can see from Fig. \ref{fig:masa}, the curvature on
the hypersurfaces $t=$ constant is concentrated along the Rindler horizon, being zero elsewhere. Even more interesting is the fact that
the boost angles are the same for all triangles on the Rindler horizon. Therefore the classical action $\widetilde{S}_G$
takes, when spacetime is flat, the form
\begin{equation} \widetilde{S}_G=\frac{1}{8\pi} A(\phi_2 - \phi_1),
\end{equation}
where $A$ is the total area of the considered part of the Rindler horizon, whereas $\phi_1$ and $\phi_2$, respectively, are the 
boost angles between the future pointing tangent vectors of the world lines of the oppositely accelerated observers at times $t_1$
and $t_2$. Therefore we may approximate the propagator as
\begin{equation} K \big( q_{ab}(t_2), t_2; q_{ab}(t_1),t_1 \big) \approx N[q_{ab}(t_1), q_{ab}(t_2)]\exp \Big[
\frac{i}{8\pi}A(\phi_2-\phi_1) \Big].
\end{equation} 

We are now prepared to enter the thermodynamics of Rindler horizons. As it is well known, thermodynamics is just quantum mechanics
in euclidean spacetime\cite{hi}. More precisely, if a system is periodic in euclidean spacetime with period $\beta$ such that after the 
elapsed euclidean time $\beta$ it returns to the original point $x$ in configuration space, then the partition function of the system
is
\begin{equation} Z(\beta) = \int dx\, K(x,-i\beta;x,0).
\end{equation}
In other words, we have replaced the lorentzian time $t$ in the propagator by the term $-i\beta$, where $\beta$ is the euclidean
time. Thermodynamically, the quantity $\beta$
may be intepreted as the inverse temperature of the system. We shall use this idea when we calculate the partition function, and
thereby the entropy, of the Rindler horizon.

\begin{figure}[!htb]
\centering
\includegraphics[scale=0.5]{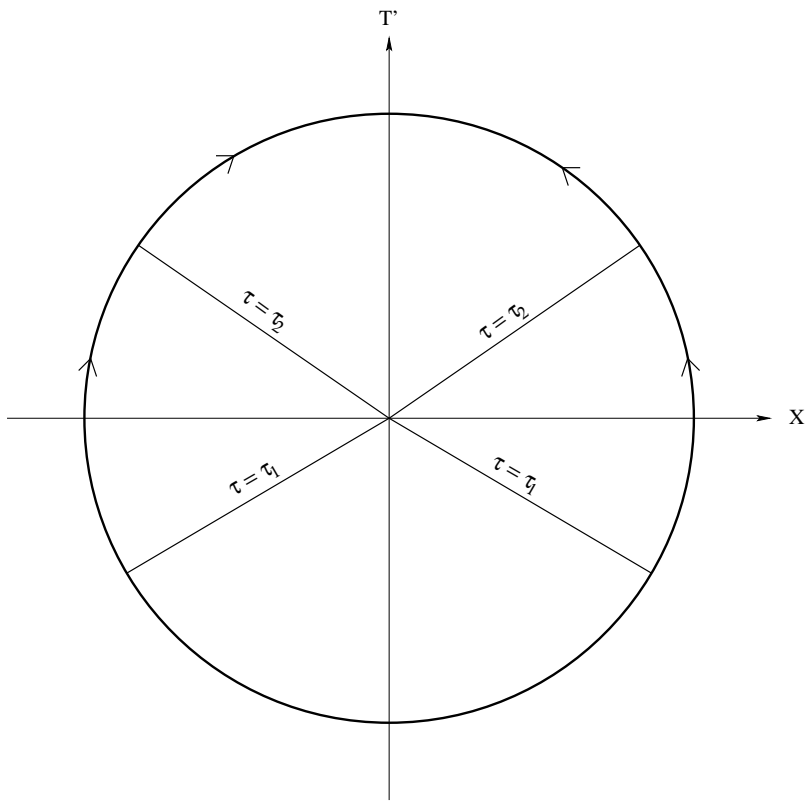}
\caption{In euclidean spacetime the world line of a uniformly accelerated observer is a circle. The hypersurfaces where
the euclidean proper time $\tau$ of the observer is constant sweep over the whole spacetime. \label{fig:pylpyra}}
\end{figure}

To begin with, consider an accelerated observer in euclidean spacetime. We just replace in Eqs. (\ref{eq:tillin}) and (\ref{eq:tallin}) 
$T$ by $-iT'$ and $t$ by $-i\tau$,
where $T'$ and $\tau$ are euclidean time coordinates. We get:
\begin{equation} T' = \frac{1}{a} \sin(a\tau),
\end{equation} 
\begin{equation} X  = \frac{1}{a} \cos(a\tau).
\end{equation}
In other words, the world line of an accelerated observer in lorentzian spacetime is, in euclidean spacetime, a \textit{circle}. In 
Fig. \ref{fig:pylpyra} we
have drawn the world lines of two accelerated observers in euclidean spacetime. The hypersurfaces $t=$ constant are replaced by the 
hypersurfaces where $\tau =$ constant. It is remarkable that spacetime geometry is now \textit{periodic} with respect to the euclidean 
time 
$\tau$: An observer returns to the same point in euclidean spacetime after the elapsed euclidean time
\begin{equation} \beta = \frac{2\pi}{a},
\end{equation}
and therefore the temperature experienced by the accelerated observer is exactly the Unruh temperature of the Rindler horizon\cite{unruh}:
\begin{equation} T_U = \frac{1}{\beta}=\frac{a}{2\pi}.
\end{equation}
In contrast to spacetime with lorentzian signature, the hypersurfaces $\tau =$ constant sweep over the whole spacetime. However, 
during just one period they sweep the spacetime two times. Because of that we must divide the action by two when we calculate 
the partition function of the Rindler horizon from the point of view of a \textit{single} observer.

We now proceed to calculate the partition function. To begin with, we consider the action in euclidean spacetime. In euclidean 
spacetime we must replace the boost angles $\phi_1$ and $\phi_2$, respectively, by the quantities $-i\theta_1$ and $-i\theta_2$, where
$\theta_1$ and $\theta_2$ are just the ''ordinary'' euclidean angles between the normals of the tetrahedra meeting at their common
two-simplex on the Rindler horizon. Hence, the euclidean propagator takes, in the semi-classical limit, the form:
\begin{equation} K_E \big( q_{ab}(\tau_2), \tau_2; q_{ab}(\tau_1),\tau_1 \big) \approx N[q_{ab}(\tau_1), q_{ab}(\tau_2)]\exp \Big[
-\frac{1}{8\pi}A(\theta_1-\theta_2) \Big].
\end{equation} 
During one period the angle between the normals of the tetrahedra meeting at this common two-simplex decreases by $4\pi$, and
therefore, taking into account that the action must be divided by two for the reasons explained before, the partition function is,
in the semi-classical limit,
\begin{equation} Z \approx \int \mathcal{D} \Big[q_{ab}\Big(\frac{2\pi}{a}\Big) \Big] N \Big[q_{ab}\Big(\frac{2\pi}{a}\Big), 
q_{ab}(0) \Big] \exp \Big(-\frac{1}{4}A \Big).
\end{equation}
Since $q_{ab}(0)= q_{ab}\big( \frac{2\pi}{a}\big)$, we integrate over a single variable $q_{ab}\big( \frac{2\pi}{a}\big)$, and
when this variable is integrated out, we get
\begin{equation} Z \approx \mathcal{N} \exp \Big( -\frac{1}{4}A \Big),
\end{equation}
where $\mathcal{N}$ is a constant.
In other words, we have obtained an exciting result that, at least in this semi-classical limit, the partition function of the
considered part of the Rindler horizon has the negative of one quarter of the area of that part in the exponential.

To calculate the entropy of the Rindler horizon we must be able to express the partition function in terms of the euclidean
period $\beta$. To this end, consider what an accelerated observer actually observes about the properties of his Rindler
horizon. Such an observer sees that, in his frame of reference, all bodies tend to fall, with a uniform acceleration $a$, towards
the Rindler horizon. According to the Principle of Equivalence the observer has no means to decide whether he is in a 
uniformly accelerating motion, or in a uniform gravitational field caused by a mass distributed uniformly along a plane. According to
Gauss' law for gravitational fields the mass enclosed by the closed two-surface $\sigma$ is given by a surface integral
\begin{equation} M = -\frac{1}{4\pi} \oint_\sigma \vec g \cdot d \vec S,
\end{equation}
where $\vec g$ is the gravitational field. If the observer is accelerated along the $x$-axis, he may conclude that he is in
the gravitational field
\begin{equation} \vec g = -a \hat{i},
\end{equation}
and therefore, because the Rindler horizon lies in the $yz$-plane, the mass contained by the part of area $A$ of the
Rindler horizon, is
\begin{equation} M = \frac{a}{4\pi}A,
\end{equation}
which follows from a consideration similar to the one performed in the standard textbooks of electromagnetism, when the charge
density of a plane is calculated by means of Gauss' law for electric fields.
In other words, the area of the considered part of the horizon may be written in terms of its observed mass $M$ and inverse
temperature $\beta$:
\begin{equation} \label{eq:tiili} A=2\beta M,
\end{equation}
and therefore the partition function $Z$ may be written, in the semi-classical limit, in terms of the average value 
$\langle M \rangle$ of the mass:
\begin{equation} Z = \mathcal{N} \exp \Big( -\frac{1}{2} \beta \langle M \rangle \Big).
\end{equation}
However, the average value of $M$ may calculated from the partition function:
\begin{equation} \langle M \rangle = -\frac{\partial}{\partial \beta} \ln Z. 
\end{equation}
This gives us a differential equation for $\langle M \rangle$:
\begin{equation} \beta \frac{\partial \langle M \rangle}{\partial \beta} = \langle M \rangle,
\end{equation}
which, in turn, has the general solution
\begin{equation} \label{eq:reilu} \langle M \rangle = C \beta,
\end{equation}
where $C$ is a constant. Hence, the partition function takes, in terms of $\beta$, the form
\begin{equation} Z(\beta) = \mathcal{N} \exp \Big( -\frac{1}{2} C \beta^2 \Big).
\end{equation}

We now proceed to calculate the entropy of the Rindler horizon. In general, the entropy of any system may be calculated from
the partition function:
\begin{equation} S = -\beta \Big( \frac{\partial}{\partial \beta} \ln Z \Big) + \ln Z.
\end{equation}
This gives us (we take $\mathcal{N} =1$):
\begin{equation} S= \frac{1}{2} C \beta^2,
\end{equation}
and it follows from Eqs. (\ref{eq:tiili}) and (\ref{eq:reilu}) that
\begin{equation} S = \frac{1}{4} A.
\end{equation}
In other words, we have obtained a result which states that, from the point of view of an accelerated observer, the entropy
of any finite part of the Rindler horizon is one quarter of the area of that part. As such our result is similar to the
Bekenstein-Hawking law for black hole entropy -- the only difference is that the area of the event horizon of the black hole has
been replaced by the area of the considered part of the Rindler horizon.

Actually, our result may be used to show that \textit{any} horizon of spacetime has an entropy which is one quarter of its
area. That is because, locally, any small enough part of any horizon may be considered, in an appropriate coordinate system,
as a part of a Rindler horizon. The entropy of the horizon as a whole, in turn, is the sum of the entropies of its parts,
and since the horizon may be constructed from tiny Rindler horizons, each having an entropy equal to one quarter of its
area, the entropy of the whole horizon is one quarter of its total area.

To see what all this means consider, as an example, the \Sw horizon. The \Sw metric in \Sw coordinates $t$ and $r$ is
\begin{equation} ds^2= -\Big(1-\frac{2M}{r} \Big) dt^2 + \frac{dr^2}{1- \frac{2M}{r}} + r^2(d\theta^2 + \sin^2 \theta\, d\phi^2), 
\end{equation}
where $M$ is the mass of the \Sw black hole. An observer at rest with respect to the coordinates $r$, $\theta$ and $\phi$
observes that the hole radiates thermal radiation with a characteristic temperature\cite{haw}
\begin{equation} \label{eq:dilli} T_H =: \frac{1}{8\pi M \sqrt{1-\frac{2M}{r}}},
\end{equation}
where we have taken into account the ''red shift'' of the temperature. In other words, an observer at rest very close
to the horizon $r=2M$ may measure an infinite temperature for the black hole radiation. 

Consider now the radiation process from the point of view of an observer in free fall, very close to the horizon.
Appropriate coordinates for the study of the properties of freely falling observers are the so-called 
\textit{Novikov coordinates}\cite{mtw} (for a detailed survey of the Novikov coordinates, see Appendix). In terms of the Novikov
coordinates ($\tau,R^*$) the \Sw line element takes the form
\begin{equation}  ds^2 = - d\tau^2 + \frac{R^{*2}+1}{R^{*2}} \Big( \frac{\partial r}
{\partial R^{*}} \Big)^2 dR^{*2} + r^2(d\theta^2 + \sin^2 \theta \, d\phi^2).
\end{equation}
For an observer in radial free fall such that $\frac{dr}{dt}=0$ when $t=0$, $R^*$ is constant and $\tau$ is the
proper time of such an observer. At the bifurcation point in the Kruskal diagram of Kruskal spacetime $R^* = \tau =0$,
and for an observer in radial free fall through the bifurcation two-surface $R^* =0$. Close to the bifurcation
two-surface we may write $\frac{\partial r}{\partial R^*}$ as a Taylor expansion (see Appendix):
\begin{equation} \label{eq:markku} \frac{\partial r}{\partial R^*} = 4MR^* + 3MR^{*3} + \frac{1}{2M}R^* \tau^2 + \dots,
\end{equation} 
and therefore the spacetime metric becomes, at the bifurcation two-surface,
\begin{equation} ds² = -d\tau^2 + (4M)^2 dR^{*2} + (2M)^2 d\Omega^2
\end{equation}
where $d\Omega^2$ is the line element on a unit two-sphere. If we rescale the Novikov coordinate $R^*$ such that we
define a new coordinate 
\begin{equation} \widetilde{R}^* := 4MR^*,
\end{equation}
we get
\begin{equation} \label{eq:qluu} ds^2 = - d\tau^2 + d\widetilde{R}^{*2} + (2M)^2 d\Omega^2.
\end{equation}
In other words, the Novikov coordinate system provides, at the bifurcation point, a coordinate system which is
locally flat, or locally inertial.

It is instructive to compare the expression (\ref{eq:qluu}) to the metric given by the Kruskal coordinates $u$ and $v$. In 
general, the \Sw line element takes, in Kruskal coordinates, the form
\begin{equation} ds^2 = \Big( \frac{32M^3}{r} \Big) e^{-r/2M} (-dv^2 +du^2) + r^2 d\Omega^2,
\end{equation}
and at the bifurcation two-surface we have
\begin{equation} ds^2 = \frac{16M^2}{e} (-dv^2 +du^2) + (2M)^2 d\Omega^2.
\end{equation}
Hence we observe that very close to the bifurcation two-surface between the Kruskal coordinates ($v,u$) and the coordinates
($\tau, \widetilde{R}^*$) there is the relationship
\begin{equation} \tau = \frac{4M}{\sqrt{e}}v,
\end{equation}
\begin{equation} \widetilde{R}^* = \frac{4M}{\sqrt{e}}u.
\end{equation}
In other words, close to the bifurcation two-surface the coordinates $\tau$ and $\widetilde{R}^*$ are, up to the rescaling
factor $4M/ \sqrt{e}$, identical to the Kruskal coordinates $v$ and $u$.

We are now coming to the similarities between the Rindler and \Sw horizons. As it is well known, the relationship between
the \Sw and Kruskal coordinates is, when $r>2M$\cite{mtw}:
\begin{equation} \label{eq:nisse} v = \Big( \frac{r}{2M}-1 \Big)^{1/2} e^{r/4M} \sinh \Big( \frac{t}{4M} \Big),
\end{equation}
\begin{equation} \label{eq:nasse} u = \Big( \frac{r}{2M}-1 \Big)^{1/2} e^{r/4M} \cosh \Big( \frac{t}{4M} \Big).
\end{equation}
To consider this transformation very close to the bifurcation two-surface, we define
\begin{equation} \delta := \sqrt{1-\frac{2M}{r}},
\end{equation}
and the transformations of Eqs. (\ref{eq:nisse}) and (\ref{eq:nasse}) become, for small $\delta$
\begin{equation} \label{eq:intti} \tau = 4M\delta \sinh \Big( \frac{\theta}{4M\delta} \Big),
\end{equation}
\begin{equation} \label{eq:antti} \widetilde{R}^* = 4M\delta \cosh \Big( \frac{\theta}{4M\delta}\Big),
\end{equation}
where we have defined a new time coordinate
\begin{equation} \theta := t\delta,
\end{equation}
which actually is the proper time of an observer at rest with respect to the \Sw coordinates. Comparing Eqs. (\ref{eq:tillin}) 
and (\ref{eq:tallin}) 
to Eqs. (\ref{eq:intti}) and (\ref{eq:antti}) we notice that observers at rest with respect to the \Sw coordinates are, 
quite simply, observers which are 
\textit{accelerated} with respect to the observers in free fall with a proper acceleration
\begin{equation} a = \frac{1}{4M\delta} = \frac{1}{4M\sqrt{1-\frac{2M}{r}}}.
\end{equation}
In other words, the appearance of the \Sw horizon from the point of view of an observer at rest with respect to the hole
is simply due to the fact that such an observer is accelerated with respect to a locally inertial observer. In this sense
there is no difference between a Rindler horizon and any sufficiently small part of a \Sw horizon. Both appear simply
because the rest frame of the observer is non-inertial. From the point of view of an inertial observer very close to the 
horizon, no horizon, nor any of its effects, such as Hawking radiation, do appear.

From this point on, the thermodynamical properties of the \Sw horizon may be obtained in exactly the same way as we 
obtained the thermodynamical properties of the Rindler horizon. For instance, the temperature of the horizon is
\begin{equation} T_H = \frac{a}{2\pi} = \frac{1}{8\pi M \sqrt{1-\frac{2M}{r}}},
\end{equation}
which is Eq. (\ref{eq:dilli}). Moreover, any sufficiently small part of the horizon has an entropy which is one 
quarter of its area, and therefore, \textit{a fortiriori}, the entropy of the hole is one quarter of its horizon area. A similar
chain of reasoning may be applied to any horizon of spacetime.

In this paper we have considered the thermodynamical properties of spacetime horizons in general. Our considerations
were based on detailed investigations of Rindler horizons. We argued, by using Regge calculus that, in the 
semi-classical limit, one may associate with any finite part of a Rindler horizon an entropy which is one quarter of
the area of that part. This is the entropy observed by an accelerated observer. Moreover, we showed that the results
originally obtained for Rindler horizons may be applied for the derivation of the thermodynamical properties of the \Sw
horizon as well. This conclusion was based on the observation that an observer at rest with respect to the horizon
is accelerated with respect to an observer in free fall through the horizon. In other words, any sufficiently
small part of a \Sw horizon may be considered as a part of a Rindler horizon. Since the entropy of the Rindler
horizon is one quarter of its area, so is the entropy of the \Sw horizon as well. A similar chain of reasoning may
be applied to any horizon, and therefore it appears that one may argue, in the semi-classical limit, that the entropy of
any spacetime horizon is, from the point of view of an observer at rest with respect to that horizon, one quarter
of its area. 

\textit{Note}: During the preparation of this paper we became aware of an interesting recent paper written by Padmanabhan\cite{pad}.
He also showed that the entropy of any horizon is one quarter of its area. His methods, however, are completely different from ours.

\begin{acknowledgements} 
We are grateful to Markku Lehto and Jorma Louko for their constructive criticism during the preparation of this paper.
\end{acknowledgements}

\appendix*
\section{Novikov Coordinates Close to the Bifurcation Point}
\label{sec:liite}
\subsection{Novikov Coordinates}
When one is considering observers in free fall towards the \Sw black hole, it is often useful to apply Novikov 
coordinates. In this coordinate system a freely-falling observer at rest with respect to the \Sw coordinates,
for the \Sw time $t=0$, remains at rest during his whole journey. At the time $t=0$ 
the observer is released in radial free fall and the \Sw radial coordinate 
$r$ of the observer can be written at the time $t=0$ as\cite{mtw} 
\begin{equation} \label{eq:R} r = r_{max} = 2M(1+R^{*2}),
\end{equation}
where $R^{*}$ is the radial coordinate in Novikov coordinates. The time coordinate in Novikov
coordinates is the proper time $\tau$ of this freely-falling observer.
Thus, in the Novikov coordinate system spacetime points 
are identified by means of the radial coordinate $R^*$ and the proper time coordinate $\tau$.

The equation of motion for an observer in free fall in \Sw spacetime is
\begin{equation}1 = \frac{C^2}{1-\frac{2M}{r}} - \frac{\dot{r}^2}{1-\frac{2M}{r}},
\end{equation}
where $C$ is a constant and the dot stands for the proper time derivative. Hence we get
\begin{equation} \label{eq:dr} \dot{r} = \pm \sqrt{\frac{2M}{r}-1+C^2}.
\end{equation}
It is easy to see that the observer is at rest with respect to the \Sw coordinate $r$ when
\begin{equation} r = r_{max} = \frac{2M}{C^2 - 1}.
\end{equation}
From equations (\ref{eq:R}) and (\ref{eq:dr}) we may solve $\dot{r}$ in terms of $R^*$ and $r$:
\begin{equation} \label{eq:dot_r} \dot{r} = \pm \Bigg[ \frac{2M}{r} - \frac{1}{R^{*2}+1} \Bigg]^{1/2}.
\end{equation} 
Furthermore, we can express the proper time $\tau$ as an integral
\begin{equation} \tau = \pm \int\limits_{2M(R^{*2}+1)}^{r} \Bigg[ \frac{2M}{r'} - \frac{1}{R^{*2}+1} 
\Bigg]^{-1/2} dr', \end{equation}
which gives the relationship between the \Sw coordinate $r$ and the Novikov coordinates $R^{*}$ 
and $\tau$:
\begin{eqnarray} \label{eq:tau} \pm \frac{\tau}{2M} &=& (1+R^{*2}) \Bigg[ \frac{r}{2M} - \frac{r^2}{4M^2(R^{*2}+1)} 
\Bigg]^{1/2} \nonumber \\ & &+ (R^{*2}+1)^{3/2} \arccos \Bigg[ \Bigg(\frac{r}{2M( R^{*2}+1)} \Bigg)^{1/2} \Bigg].
\end{eqnarray}

The Schwarzschild line element can also be written in terms of the Novikov coordinates $R^{*}$ and $\tau$, 
and it takes the form
\begin{equation} \label{eq:le} ds^2 = - d\tau^2 + \frac{R^{*2}+1}{R^{*2}} \Bigg( \frac{\partial r}
{\partial R^{*}} \Bigg)^2 dR^{*2} + r^2(d\theta^2 + \sin^2 \theta \, d\phi^2).
\end{equation}
Here $r$ should be thought as a function $r(\tau,R^{*})$ given implicitly by Eq. (\ref{eq:tau}). The point 
$\tau = R^{*} = 0$ is the bifurcation point $u = v = 0$ of Kruskal spacetime. At this point $r = 2M$. 
It is possible to obtain the Taylor expansion for $\big( \frac{\partial r}{\partial R^{*}} \big)$ and 
thereby write the line element (\ref{eq:le}) close to the bifurcation point in terms of $R^*$ and $\tau$.
 
\subsection{Taylor Expansion for $\frac{\partial r}{\partial R^*}$}
Let us write Eq. (\ref{eq:tau}) in the form
\begin{equation} \label{eq:f} \pm \frac{\tau}{2M}= f\big( R^*,r(R^*,\tau) \big),
\end{equation}
where $f\big( R^*,r(R^*,\tau) \big)$ is the function at the right hand side of Eq. (\ref{eq:tau}). When one differentiates
both sides of Eq. (\ref{eq:f}) with respect to $R^*$, one gets the result
\begin{equation} \label{eq:nolla} 0 = \frac{d}{dR^*} f\big( R^*,r(R^*,\tau)\big) = \frac{\partial f}{\partial R^*}+
\frac{\partial f}{\partial r} \frac{\partial r}{\partial R^*}.
\end{equation}
It is easy to show that
\begin{eqnarray} \frac{\partial f}{\partial R^*} &=& 2R^* \Bigg[\frac{r}{2M}-
\frac{r^2}{4M^2 (R^{*2}+1)} \Bigg]^{1/2} 
+ \frac{R^* r}{2M} \Bigg[\frac{r}{2M}- \frac{r^2}{4M^2(R^{*2}+1)}\Bigg]^{-1/2} \nonumber \\
& & + \frac{R^* r^2}{4M^2(R^{*2}+1)} \Bigg[\frac{r}{2M}- \frac{r^2}{4M^2(R^{*2}+1)} \Bigg]^{-1/2} \nonumber \\
& & + 3R^*(R^{*2} +1)^{1/2} \arccos \Bigg[ \Bigg( \frac{r}{2M(R^{*2}+1)} \Bigg)^{1/2} \Bigg]
\end{eqnarray}
and
\begin{eqnarray} \frac{\partial f}{\partial r} &=& 
\frac{R^{*2}+1}{2r} \Bigg[\frac{r}{2M}- \frac{r^2}{4M^2(R^{*2}+1)} \Bigg]^{1/2} \nonumber \\ 
& & -\frac{R^{*2}+1}{4M} \Bigg[\frac{r}{2M}- \frac{r^2}{4M^2(R^{*2}+1)} \Bigg]^{-1/2}.
\end{eqnarray}
Thus, according to Eq. (\ref{eq:nolla}), we get
\begin{eqnarray} \label{eq:rR} \frac{\partial r}{\partial R^*} &=&
-\frac{\partial f / \partial R^*}{\partial f / \partial r} =
6MR^* -\frac{R^* r}{R^{*2}+1} \nonumber \\ & &+ 12M^2 R^*\Bigg[\frac{(R^{*2} +1)}{2Mr}- 
\frac{1}{4M^2} \Bigg]^{1/2} \arccos \Bigg[ \Bigg( \frac{r}{2M(R^{*2}+1)} \Bigg)^{1/2} \Bigg].
\end{eqnarray}

Instead of obtaining the Taylor expansion directly from Eq. (\ref{eq:rR}), one may define 
\begin{equation} \label{eq:u} w:=R^{*2},
\end{equation}
\begin{equation} \label{eq:g_f} g\big( w,r(w,\tau) \big) := \frac{1}{R^*} \frac{\partial r}{\partial R^*}
\end{equation}
and try to calculate the Taylor expansion 
\begin{eqnarray} \label{eq:g} g(w,r) &=& g \Big\arrowvert_{w=0} \ +\ \frac{\partial g}{\partial w} 
\Bigg\arrowvert_{w=0} w \ + \ \frac{\partial g}{\partial \tau}\Bigg\arrowvert_{w=0} \tau \ + \ \frac{1}{2!}
\frac{\partial^2 g}{\partial w^2}\Bigg\arrowvert_{w=0} w^2 \nonumber \\
& &+\ \frac{\partial^2 g}{\partial \tau \partial w}\Bigg\arrowvert_{w=0} w\tau \ 
+ \ \frac{1}{2!}\frac{\partial^2 g}{\partial \tau^2}\Bigg\arrowvert_{w=0} \tau^2 \ +  \dots
\end{eqnarray}
with simpler expressions. If one needs at most third-order terms of the Taylor expansion for 
$\frac{\partial r}{\partial R^*}$, one has to calculate only the three first terms and the last term 
in Eq. (\ref{eq:g}).

Putting $w=0$ and $r=2M$ one finds that
\begin{equation} \label{eq:t1} g \Big\arrowvert_{w=0} = 4M.
\end{equation}
When obtaining the next term for Eq. (\ref{eq:g}) one has to use the chain rule: 
\begin{equation} \label{eq:gu} \frac{\partial g(w,\tau)}{\partial w} = \frac{\partial g(w,r)}{\partial w} +
\frac{\partial g(w,r)}{\partial r} \frac{\partial r}{\partial w}.
\end{equation}
It should be noted that on the left hand side of Eq. (\ref{eq:gu}) $g$ is considered as a function of $w$ and
$\tau$, whereas on the right hand side $g$ is considered as a function of $w$ and $r$.
It is straightforward to show that
\begin{eqnarray} \label{eq:pilli} \frac{\partial g(w,r)}{\partial w} &=& \frac{r}{(w+1)^2} + \frac{3M}{w+1} \nonumber \\ 
& & +\frac{3M}{r} \Bigg( \frac{w+1}{2Mr}-\frac{1}{4M^2}\Bigg)^{-1/2} \arccos \Bigg[ \Bigg( \frac{r}{2M(w+1)} 
\Bigg)^{1/2} \Bigg]
\end{eqnarray} 
and 
\begin{eqnarray}\label{eq:pulla} \frac{\partial g(w,r)}{\partial r} &=& -\frac{1}{w+1} - \frac{3M}{r} \nonumber \\
& & -\frac{3M(w+1)}{r^2} \Bigg( \frac{w+1}{2Mr}-\frac{1}{4M^2}\Bigg)^{-1/2} \arccos \Bigg[ \Bigg( \frac{r}{2M(w+1)} 
\Bigg)^{1/2} \Bigg].
\end{eqnarray} 
When solving $\frac{\partial r}{\partial w}$ one has to use a similar method that was used in solving 
$\frac{\partial r}{\partial R^*}$. Let us write Eq. (\ref{eq:tau}) in the form
\begin{equation} \label{eq:h} \pm \frac{\tau}{2M}= h\big( w,r(w,\tau)\big).
\end{equation}
When one differentiates both sides of this equation with respect to $w$, one gets
\begin{equation} \label{eq:nolla2} 0 = \frac{\partial h}{\partial w}+
\frac{\partial h}{\partial r} \frac{\partial r}{\partial w}.
\end{equation}
It can be showed that 
\begin{eqnarray} \frac{\partial h}{\partial w} &=& \Bigg[\frac{r}{2M}-
\frac{r^2}{4M^2 (w+1)} \Bigg]^{1/2} 
+ \frac{r}{4M} \Bigg[\frac{r}{2M}- \frac{r^2}{4M^2(w+1)}\Bigg]^{-1/2} \nonumber \\
& & + \frac{r^2}{8M^2(w+1)} \Bigg[\frac{r}{2M}- \frac{r^2}{4M^2(w+1)} \Bigg]^{-1/2} \nonumber \\
& & + \frac{3}{2}(w +1)^{1/2} \arccos \Bigg[ \Bigg( \frac{r}{2M(w+1)} \Bigg)^{1/2} \Bigg],
\end{eqnarray}
and
\begin{eqnarray} \frac{\partial f}{\partial r} &=& 
\frac{w+1}{2r} \Bigg[\frac{r}{2M}- \frac{r^2}{4M^2(w+1)} \Bigg]^{1/2} \nonumber \\ 
& & - \frac{w+1}{4M} \Bigg[\frac{r}{2M}- \frac{r^2}{4M^2(w+1)} \Bigg]^{-1/2},
\end{eqnarray}
which, together with Eq. (\ref{eq:nolla2}), lead to
\begin{eqnarray} \label{eq:ru} \frac{\partial r}{\partial w} &=&
-\frac{\partial h / \partial w}{\partial h / \partial r} =
3M -\frac{r}{2(w+1)} \nonumber \\ & & + 6M^2 \Bigg(\frac{w+1}{2Mr}- 
\frac{1}{4M^2} \Bigg)^{1/2} \arccos \Bigg[ \Bigg( \frac{r}{2M(w+1)} \Bigg)^{1/2} \Bigg].
\end{eqnarray}
Now, according to equations (\ref{eq:gu}), (\ref{eq:pilli}), (\ref{eq:pulla}) and (\ref{eq:ru}) we get
\begin{eqnarray} \frac{\partial g(w,\tau)}{\partial w} &=& \frac{3r}{2(w+1)^2} + \frac{9M}{2(w+1)}
- \frac{9M^2}{r} \nonumber \\ & &
- \frac{18M^3 (w+1)}{r^2} \arccos^2 \Bigg[ \Bigg( \frac{r}{2M(w+1)} \Bigg)^{1/2} \Bigg] \nonumber \\
& & -\frac{6M^2}{w+1}  \Bigg(\frac{w+1}{2Mr}- \frac{1}{4M^2}\Bigg)^{1/2}
\arccos \Bigg[ \Bigg( \frac{r}{2M(w+1)} \Bigg)^{1/2} \Bigg] \nonumber \\ 
& & - \frac{36M^3}{r} \Bigg(\frac{w+1}{2Mr}- \frac{1}{4M^2}\Bigg)^{1/2}
\arccos \Bigg[ \Bigg( \frac{r}{2M(w+1)} \Bigg)^{1/2} \Bigg].
\end{eqnarray}
Again, putting $w=0$ and $r=2M$ one finds that
\begin{equation} \label{eq:t2} \frac{\partial g}{\partial w} \Bigg\arrowvert_{w=0} = 3M.
\end{equation}

To determine the next term in Eq. (\ref{eq:g}) we must solve $\frac{\partial g}{\partial \tau}$. The chain rule,
together with equations (\ref{eq:dot_r}) and (\ref{eq:pulla}), gives
\begin{eqnarray} \frac{\partial g(w,\tau)}{\partial \tau} &=& \frac{\partial g(w,r)}{\partial r}
\frac{\partial r}{\partial \tau} = \frac{\partial g}{\partial r}
\Bigg(\frac{2M}{r}- \frac{1}{w+1} \Bigg)^{1/2} \nonumber \\
&=& -\Bigg[\frac{2M}{(w+1)^{3/2}} + \frac{6M^2}{r(w+1)^{1/2}} \Bigg]
\Bigg(\frac{w+1}{2Mr}- \frac{1}{4M^2} \Bigg)^{1/2} \nonumber \\
& & -\frac{6M^2(w+1)^{1/2}}{r^2} \arccos \Bigg[ \Bigg( \frac{r}{2M(w+1)} \Bigg)^{1/2} \Bigg]. 
\end{eqnarray} 
Thus, we get
\begin{equation} \label{eq:t3} \frac{\partial g}{\partial \tau} \Bigg\arrowvert_{w=0} = 0.
\end{equation}

Let us now define $g_1 := \frac{\partial g(w,\tau)}{\partial \tau}$. Again, using the chain rule and Eq. (\ref{eq:dot_r}) 
we get
\begin{equation} \label{eq:kille} \frac{\partial^2 g(w,\tau)}{\partial \tau^2} = 
\frac{\partial g_1}{\partial \tau} = \frac{\partial g_1}{\partial r} \frac{\partial r}{\partial \tau}
=\frac{\partial g_1}{\partial r} \Bigg(\frac{2M}{r}- \frac{1}{w+1} \Bigg)^{1/2}.
\end{equation}
One can show that
\begin{eqnarray} \frac{\partial g_1}{\partial r} &=& \Bigg[ \frac{1}{2r^2 (w+1)^{1/2}} +
\frac{3M(w+1)^{1/2}}{r^3} \Bigg] \Bigg(\frac{w+1}{2Mr}- \frac{1}{4M^2}\Bigg)^{-1/2} \nonumber \\
& & + \frac{6M^2}{r^2(w+1)^{1/2}} \Bigg(\frac{w+1}{2Mr}- \frac{1}{4M^2}\Bigg)^{1/2} \nonumber \\
& & + \frac{12M^2 (w+1)^{1/2}}{r^3}\arccos \Bigg[ \Bigg( \frac{r}{2M(w+1)} \Bigg)^{1/2} \Bigg], 
\end{eqnarray}
which, together with Eq. (\ref{eq:kille}), leads to
\begin{eqnarray}  \frac{\partial^2 g(w,\tau)}{\partial \tau^2} &=& \frac{M}{r^2(w+1)}+\frac{6M^2}{r^3}
+ \frac{12M^3}{r^2(w+1)} \Bigg(\frac{w+1}{2Mr}- \frac{1}{4M^2}\Bigg) \nonumber \\
& & + \frac{24M^3}{r^3} \Bigg(\frac{w+1}{2Mr}- \frac{1}{4M^2}\Bigg)^{1/2} 
\arccos \Bigg[ \Bigg( \frac{r}{2M(w+1)} \Bigg)^{1/2} \Bigg].
\end{eqnarray}
Putting again $w=0$ and $r=2M$ gives
\begin{equation} \label{eq:t4} \frac{\partial^2 g}{\partial \tau^2} \Bigg\arrowvert_{w=0} = \frac{1}{M}.
\end{equation}

Finally, using equations (\ref{eq:t1}), (\ref{eq:t2}), (\ref{eq:t3}) and (\ref{eq:t4}) we may 
write down the Taylor expansion for $\frac{\partial g}{\partial w}$. Furthermore, by using definitions 
(\ref{eq:u}) and (\ref{eq:g_f}) we may obtain the Taylor expansion for $\frac{\partial r}{\partial R^*}$.
We get:
\begin{equation} \frac{\partial r}{\partial R^*} = 4MR^*+3MR^{*3}+\frac{1}{2M}R^* \tau^2 + \dots,
\end{equation}
which is Eq. (\ref{eq:markku}).

\end{document}